\documentclass[twocolumn,english]{article}

\usepackage[utf8]{inputenc}

\usepackage[]{babel}
\usepackage[big]{dgruyter}
\usepackage{microtype}
\usepackage{standalone}
\usepackage{siunitx}
\usepackage{tikz}
\usepackage{pgfplots}
\usepackage{pgfplotstable}
\usepackage{csquotes}
\usepackage[backend=biber,sorting=none,style=numeric-comp]{biblatex}
\bibliography{literatur.bib}
\pgfplotsset{compat=newest}
\usetikzlibrary{external}
\tikzsetexternalprefix{tikz/}
\usepgfplotslibrary{groupplots}
\usepackage{wrapfig}

\AtBeginEnvironment{appendices}{\crefalias{section}{appendix}}
\usepackage[acronym,nomain]{glossaries}

\usepackage{listings}
\usepackage{xcolor}

\definecolor{codegreen}{rgb}{0,0.6,0}
\definecolor{codegray}{rgb}{0.5,0.5,0.5}
\definecolor{codepurple}{rgb}{0.58,0,0.82}
\definecolor{backcolour}{rgb}{0.95,0.95,0.92}

\lstdefinestyle{mystyle}{
	backgroundcolor=\color{backcolour},   
	commentstyle=\color{codegreen},
	keywordstyle=\color{magenta},
	stringstyle=\color{codepurple},
	basicstyle=\footnotesize,
	breakatwhitespace=false,         
	breaklines=true,                 
	captionpos=b,                    
	keepspaces=true,                 
	numbers=left,                    
	numbersep=5pt,                  
	showspaces=false,                
	showstringspaces=false,
	showtabs=false,                  
	framextopmargin=0ex,
	framexbottommargin=0ex,
	framexleftmargin=0em,
	framexrightmargin=0em,
	numbers=none,
	morekeywords = [1]{AlgebraicDifferentiator},
}
\usepgfplotslibrary{groupplots,fillbetween}
\usepackage[capitalise]{cleveref}

\crefname{listing}{Code snippet}{Code snippets}
\Crefname{listing}{Code snippet}{Code snippets}

\usepackage{array}
\usepackage{xcolor}
\newcolumntype{L}[1]{>{\raggedright\let\newline\\\arraybackslash\hspace{0pt}}m{#1}}
\newcolumntype{C}[1]{>{\centering\let\newline\\\arraybackslash\hspace{0pt}}m{#1}}
\newcolumntype{R}[1]{>{\raggedleft\let\newline\\\arraybackslash\hspace{0pt}}m{#1}}
\usepackage{multirow}

\newcommand{\MATLAB}{\textsc{Matlab}{}}
\usepackage{color, colortbl}

\newcommand{\JacN}{N}
\newcommand{\Jacalpha}{\alpha}
\newcommand{\Jacbeta}{\beta}

\newcommand{\JacT}{T}
\newcommand{\AlgTdis}{n_{\mathrm{s}}}
\newcommand{\JacDeltat}{\delta_{t}}
\newcommand{\JacDiff}{g^{(\alpha,\beta)}_{N,T,\vartheta}}
\newcommand{\JacContiKernel}[1]{\JacDiff(#1)}

\newcommand{\wAlphaBeta}{w^{(\Jacalpha, \Jacbeta)}}
\newcommand{\JacTheta}{\vartheta}
\newcommand{\JacGamma}{\Gamma}

\newcommand{\SkalarProdukt}[1]{\langle #1\rangle}

\newcommand{\setR}{\mathbb{R}}
\newcommand{\setN}{\mathbb{N}}
\newcommand{\samplingTime}{t_{\text{s}}}

\newcommand{\norm}[1]{ \left\lVert#1\right\rVert } 
\newcommand{\abs}[1]{ \left|#1\right| } 	
\newcommand{\diffd}{ \mathrm{d} }
\newcommand{\jacPol}{\mathrm{P}}
\newcommand{\stepJac}{h_{N,T,\vartheta}^{(\alpha,\beta)}}
\newcommand{\AlgDiffF}{\mathcal{G}^{(\alpha,\beta)}_{N,T,\vartheta}}
\newcommand{\AlgDiffFTilde}{\tilde{\mathcal{G}}^{(\alpha,\beta)}_{N,T,\vartheta}}
\newcommand{\imag}{\iota}

\newcommand{\regBetaFcn}[1]{I_{#1}}
\newcommand{{\BetaFcn}}[2]{\mathrm{B}\!\left(#1,#2\right)}
\newcommand{\diff}{\mathrm{d}}
\newcommand{\AlgOmegaC}{\omega_{\text{c}}}
\newacronym{FIR}{FIR}{finite impulse response}

\theoremstyle{dgdef}

\newtheoremstyle{break}
{\topsep}{\topsep}%
{\itshape}{}%
{\bfseries}{}%
{\newline}{}%
\theoremstyle{break}
\newtheorem{exmp}{Example}[]

\usepackage[toc,page]{appendix}

\begin{document}

  \articletype{Tools}

  \author*[1]{Amine Othmane}
  \author[2]{Joachim Rudolph}
  \runningauthor{Amine Othmane}
  \affil[1]{Systems Modeling and Simulation, Saarland University, Campus A5 1, Saarbrücken, 66123, Germany, e-mail: amine.othmane@uni-saarland.de}
  \affil[2]{Chair of Systems Theory and Control Engineering, Saarland University, Campus A5 1, Saarbrücken, 66123, Germany, e-mail: j.rudolph@lsr.uni-saarland.de}
  \title{AlgDiff: An open source toolbox for the design, analysis and discretisation of algebraic differentiators}  
  \runningtitle{Designing algebraic differentiators using AlgDiff}
  \subtitle{AlgDiff: Eine Open-Source-Toolbox für den Entwurf, die Analyse und die Diskretisierung algebraischer Ableitungsschätzer}
  
  \abstract{Algebraic differentiators have attracted much interest in recent years. Their simple implementation as classical finite impulse response digital filters and systematic tuning guidelines may help to solve challenging problems, including, but not limited to, nonlinear feedback control, model-free control, and fault diagnosis. This contribution introduces the open source toolbox AlgDiff for the design, analysis and discretisation of algebraic differentiators.} 
  
  \transabstract[ngerman]{Algebraische Ableitungsschätzer haben in den letzten Jahren großes Interesse erlangt. Dank ihrer einfachen Implementierung als klassische digitale Filter mit endlicher Impulsantwort und der systematischen Parametrierungsansätze können sie zur Lösung anspruchsvoller regelungstechnischer Probleme beitragen. Dieser Beitrag stellt die Open-Source-Toolbox AlgDiff für den Entwurf, die Analyse und die Diskretisierung algebraischer Ableitungsschätzer vor.}
  \transkeywords[ngerman]{Numerisches Differenzieren, orthogonale Polynome, lineare Filterung}
  \keywords{Numerical differentiation, orthogonal polynomials, linear filtering}

\maketitle

\section{Introduction}
Numerical estimation of derivatives of measured signals is a long-standing and challenging problem because small perturbations in the measurements may yield significant errors in the estimates. Numerous approaches have been proposed to address this problem: frequency-domain digital filter design \cite{chen1995,rader2006}, Tikhonov regularisation \cite{murio2011,cullum1971}, observers design \cite{levant1998,dabroom1997,dabroom1999,levant2003,chitour2002}, local least-squares fitting of data by polynomials \cite{Savitzky1964,Madden1978},  and differentiation by integration methods \cite{diekema2012,cioranescu1934,lanczos1988}. Algebraic differentiators have been proposed in \cite{mboup2007b,mboup2009a} and further discussed and extended in \cite{reger2009,reger2008,liu2010,liu2011b,kiltz2013,kiltz2017,mboup2014,mboup2018,othmane2021ejc}, for example. A detailed introduction to these differentiation techniques, their historical evolution, and links to established methods in the literature are summarised in the free-access survey \cite{othmane2021b}. 
\begin{figure*}[h!]
	\centering
	\includegraphics[scale=0.275]{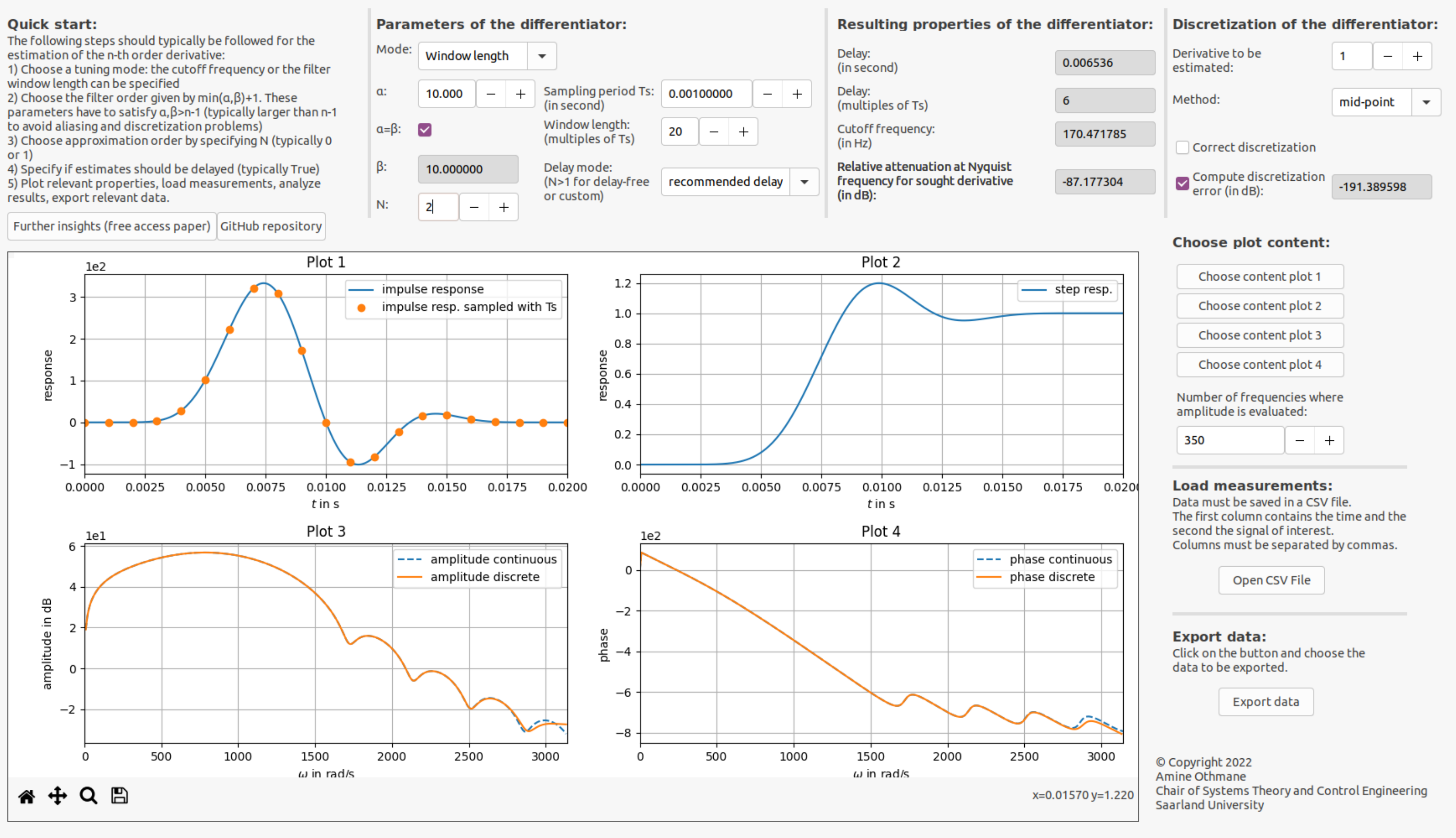}
	\caption{Screenshot of the GUI for a specific parametrisation of algebraic differentiators.}
	\label{fig:gui}
\end{figure*}

Algebraic differentiators may be interpreted as differentiation by integration methods and may be implemented as classical \acrfull{FIR} digital filters. The design of these differentiators involves the judicious choice of up to five parameters, which can be challenging. For example, it has been pointed out in \cite{mboup2007b,mboup2009a} that accepting a slight but known delay in the estimates reduces the order of the approximation error. In \cite{mboup2014}, it has been observed that delay-free differentiators may exhibit intolerable gains in their amplitude spectra. 

Analysing the effects of the parameters on the frequency-domain properties has been very useful in deriving systematic tuning guidelines. In \cite{kiltz2013}, it has been shown that it is possible to approximate the differentiators as low-pass filters. The tuning parameters can then be determined by specifying the desired cutoff frequency and stopband slope. Further effects of the parameters on the frequency-domain properties have been discussed in \cite{kiltz2017,mboup2018,othmane2021ejc}, for example. The simple implementation of these differentiators and their good robustness with respect to measurement disturbances has motivated their wide use. An extensive list of applications including but not limited to parameter estimation, fault detection and feedback control has been published in \cite{othmane2021b}.

A first version of an open-source toolbox implementing all necessary functions for the design, analysis, and discretisation of algebraic differentiators has been made available with the survey \cite{othmane2021b}. The toolbox is implemented in Python. Numerous examples show how it can be used in \MATLAB{}. The toolbox \cite{othmane2022Code} allows users to specify desired frequency-domain properties and estimating derivatives of signals in very few lines of code, for example. Time-domain and frequency-domain properties are automatically calculated and can be easily accessed. In the current contribution a step-by-step tutorial is provided for algebraic differentiators using this toolbox. Relevant properties are recalled and demonstrated. Discretisation effects and a special parametrisation for the exact annihilation of harmonic signals first discussed in \cite{kiltz2013} are also addressed using simple examples. The source code of the latter examples is made freely available in the toolbox repository for Python and \MATLAB{}.

This tutorial-like article is structured as follows. \cref{sec:toolbox} describes the main features of the toolbox. \cref{sec:background} recalls the basics of algebraic differentiators required for the remaining parts. \cref{sec:properties_alg_diff} constitutes the main part of this work and recalls the properties of algebraic differentiators in the light of the toolbox. Example code snippets are used to show how the different functions can be used. A short conclusion is provided in Section \ref{sec:conclusion}. \cref{sec:app1,sec:app2,sec:app3,sec:app4,sec:app5} summarise explicit mathematical expressions related to algebraic differentiators that are provided for the sake of completeness.

\section{The toolbox AlgDiff}

\label{sec:toolbox}
The toolbox contains an implementation in Python for all necessary functions for the design, analysis, and discretisation of algebraic differentiators. The toolbox is available under the BSD-3-Clause License, making it suitable for academic and industrial/commercial use. An interface to \MATLAB{} is included in the package, and several use cases are documented. The repository also contains detailed documentation of the Python class and its methods. The prerequisites for using the toolbox are described in detail.

\begin{table*}[h!]
	\caption{Extract from \cite{othmane2021b}: Interpretations of $g=\JacDiff$ and the practical usage of each.}
	\centering
	\renewcommand{\arraystretch}{1.5}
	\begin{tabular}{L{2.9cm}L{5.45cm}L{5.5cm}} 
		Context & Interpretation  & Practical usage \\ \midrule
		Approximation-theoretic&Polynomial approximation of derivative using a Hilbert reproducing kernel &Analysis of estimation properties and  relation to established differentiation methods\\\midrule 
		\multirow{2}{*}[-23pt]{Systems-theoretic}&Linear time-invariant filtering of the sought derivative: 
		\begin{center}
			\includegraphics{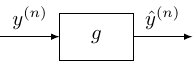}
		\end{center}
		&Analysis of filter properties and parametrisation\\
		&Linear time-invariant filtering of the measured signal:
		\begin{center}
			\includegraphics{figures/signal1.pdf}
		\end{center} 
		&Discretisation and implementation\\
	\end{tabular}
	\label{tab:summaryAlgDiff}
\end{table*}

The repository also includes a graphical user interface (GUI) that further simplifies the design of algebraic differentiators. The GUI enables specifying desired time-domain and frequency-domain properties for the filters. Relevant properties like the approximation delay,  the discretisation error,  the cutoff frequency, and  the filter window length are displayed. Numerical values for  amplitude spectra, step responses, impulse responses, and filter coefficients can be exported. Different approaches can be used to discretise the differentiators. Measurement data can be loaded into the GUI and derivatives can then be easily computed.  \cref{fig:gui} shows a screenshot of the GUI for a specific algebraic differentiator.
\section{Background material on algebraic differentiators}
\label{sec:background}
This section provides a brief summary of the fundamentals of algebraic differentiators, which were originally developed in \cite{mboup2007b,mboup2009a}. A survey on the existing results, interpretations, relationships to established methods, tuning guidelines, and applications can be found in \cite{othmane2021b}.

\subsection{Derivative estimation}
\begin{figure*}[h!]
	\centering
	\includegraphics{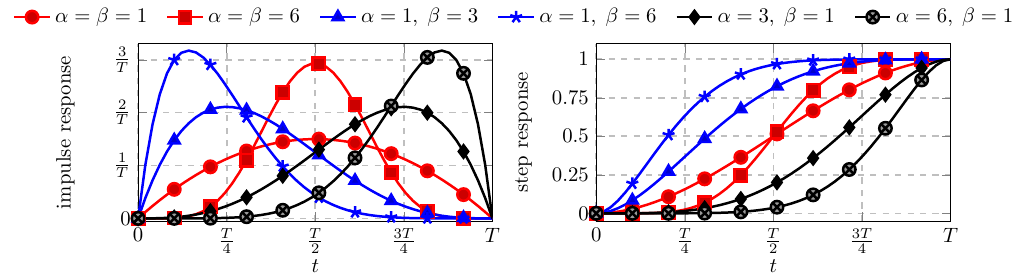}
	\caption[Impulse and step responses of Jacobi diffs. for $N=0$]{Impulse and step responses of algebraic differentiators with parameters $\alpha,\beta\in\{1,3,6\}$, and $N=0$.}
	\label{fig:stepResponse0}
\end{figure*}
While the early works \cite{mboup2007b,mboup2009a} use differential algebraic manipulations and operational calculus to derive the differentiators, later ones such as \cite{liu2011b,kiltz2013,kiltz2017,othmane2021ejc,othmane2022thesis} use time-domain derivations. This shall be briefly recalled.

Let ${\mathcal{I}_t=[t-T,t]\subset\setR}$, $T>0$, be an arbitrary closed interval and consider a Lebesgue integrable signal $y: \mathcal{I}_t\rightarrow \setR$.  The derivatives up to a finite order $n<\min\{\alpha,\beta\}+1$, with $\alpha,\beta\in\mathbb{R}$ and $\alpha,\beta>-1$,  of $y$ can be approximated using truncated generalised Fourier expansions of order $N$ based on orthogonal Jacobi polynomials as 
\begin{align}
\hat{y}^{(n)}(t) & = \int_{t-\JacT}^{t}g^{(n)}(t-\tau)y(\tau)\diffd\tau,\; g(\tau)=\JacContiKernel{\tau},
\label{eq:JacDiffApp}
\end{align}
with $\JacDiff$ defined in \eqref{eq:JacDiffKernel}. In the following, the kernel $\JacDiff$ and the parameter $T$ are called algebraic differentiator and window length, respectively. The scalars $\alpha$ and $\beta$ parametrise the weight function of the Jacobi polynomials. \cref{sec:interpretations} discusses their effects on the filter properties.

The estimate $\hat{y}^{(n)}$ corresponds to a delayed approximation of $y^{(n)}$ at time $t$ and the delay, denoted by $\JacDeltat$, can be parametrised by $\JacTheta$.
It can be expressed as
\begin{align*}
\JacDeltat&=\begin{cases}
\tfrac{\alpha+1}{\alpha+\beta+2}T,&N=0,\\
\tfrac{1-\vartheta}{2}T,&N\neq0.
\end{cases}
\end{align*}
If $\vartheta$ is a zero of the polynomial $\jacPol_{N+1}^{(\Jacalpha, \Jacbeta)}$, which yields a small \emph{but known} delay, the order of the approximation is increased, as first pointed out in \cite{mboup2009a} (see also \cite{kiltz2017,othmane2021b}). Alternatively, a delay-free approximation can be achieved for $\vartheta=1$ or a prediction for $\JacDeltat<0$ (see also \cite{kiltz2012,kiltz2017,othmane2021b}).

\subsection{Interpretations}
\label{sec:interpretations}

As discussed in \cite{othmane2021b}, different interpretations can be attached to algebraic differentiators. Two important ones are recalled in \cref{tab:summaryAlgDiff}. The approximation-theoretic interpretation stems from the time-domain derivation of the differentiators. It is helpful for the analysis of the approximation error and delays. The systems-theoretic interpretation is a direct consequence of the definition of the approximated derivative in \eqref{eq:JacDiffApp}: The estimate is the output of a linear time invariant filter where the kernel is the $n$-th order derivative of $\JacDiff$.  Applying an integration by parts in \eqref{eq:JacDiffApp} shows that the estimate can be interpreted as the output of a filter with kernel $\JacDiff$. The filter output, then, is the sought derivative. This interpretation will be used in the next section to design differentiators using the toolbox.

\section{Algebraic differentiators in the light of AlgDiff}
This section reviews the properties of algebraic differentiators in the time and frequency domains and shows how the toolbox can be used to analyse these filters. First, the continuous-time differentiators are considered. Then, discretisation issues are discussed and functions of the toolbox that can be used to ensure that the discrete-time filter preserves the properties of the continuous-time ones are introduced. Finally, the estimation of the derivative of an example signal is considered. The code snippets provided in this section are written in Python to adhere to the open-source philosophy of the package. However, the repository contains the corresponding \MATLAB{} code.
\label{sec:properties_alg_diff}
\subsection{Impulse and step responses}
The impulse and step responses of an algebraic differentiator $\JacDiff$ have been derived in \cite{kiltz2017} and are summarised in \cref{sec:app:responses}.  \cref{ex:impulse} shows how the toolbox can be used to easily compute impulse responses, their derivatives and step responses of an algebraic differentiator.

\begin{figure*}[h!]
	\centering
	\includegraphics{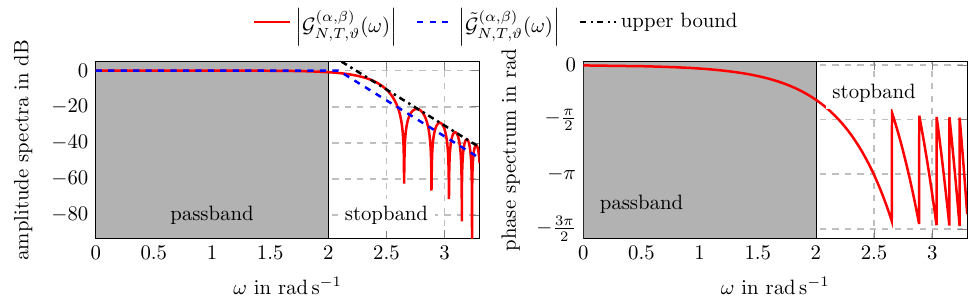}
	\caption{Amplitude and phase spectra of the differentiator from \cref{list:frequency-domain}. The approximation \eqref{eq:lowPassFilterApp} and the upper bound from \protect\cite{kiltz2017,othmane2021ejc} are also given.}
	\label{fig:spectra}
\end{figure*}

\begin{exmp}[Impulse and step responses]
	\label{ex:impulse}
	The impulse and step responses of an algebraic differentiator with parameter $\alpha=4$, $\beta=4$, $N=0$, and $T=0.1$ can be computed using the toolbox as shown in \cref{list:responses}.
	\begin{center}
		\begin{minipage}{\linewidth}
\begin{lstlisting}[escapeinside={(*}{*)},frame=single,caption={Evaluation of the impulse and step responses of an algebraic differentiator.}, label=list:responses,nolol]
# Create an instance of the class
diffA = AlgebraicDifferentiator(N=0,alpha=4.,beta=4.,T=0.1)
# Evaluate the impulse and step responses for a given time range (*${\color{codegreen}t\in[-0.05,\;0.05]}$*)
t = np.linspace(-0.05,0.1,100)
# Order of derivative of impulse res.
order_der = 0
# Evaluate the impulse response
g = diffA.evalKernelDer(order_der,t)
# Evaluate the step response
h = diffA.get_stepResponse(t)\end{lstlisting}
\end{minipage}
		\end{center}
	
\cref{fig:stepResponse0} shows the impulse and step responses of algebraic differentiators for $N=0$ and different values of $\alpha$ and $\beta$. It can be observed that increasing $\beta$ and keeping $\alpha$ constant causes a stronger weighting of the impulse response towards 0. This increase implies a stronger weighting of more recent values of the sought derivative and corresponds to the reduction of the estimation delay. In contrast, increasing $\alpha$ and keeping $\beta$ constant yields a stronger weighting of the impulse response in the direction of $T$, which leads to a stronger weighting of older values of the sought derivative and, thus, to an increase of the estimation delay. The effects of the parameters on the delay has been analysed analytically in \cite{liu2011b,mboup2009a,othmane2021ejc,othmane2021b}, for example.
\qedsymbol
\end{exmp}
The analytical properties of the impulse response and its derivatives can be used to derive tuning guidelines for fault detection algorithms, for example. Approaches have been developed and experimentally validated in \cite{kiltz2013,kiltz2014,kiltz2017,othmane2022thesis}, for example.

\subsection{Frequency-domain properties}
By recalling that the estimate of the $n$-th order derivative of $y$ is a convolution product of the signal and the $n$-th order derivative of the differentiator $\JacDiff$, it follows that
\begin{equation*}
\mathcal{F}\left\{\hat{y}^{(n)}\right\}(\omega)=(\imag\omega)^n\AlgDiffF(\imag\omega)\mathcal{Y}(\omega), \quad \imag^2=-1,
\end{equation*}
with $\mathcal{F}\left\{\hat{y}^{(n)}\right\}$, $\AlgDiffF$, and  $\mathcal{Y}$  the Fourier transforms of $\hat{y}^{(n)}$, $\JacDiff$, and $y$, respectively. For the sake of completeness, the closed form of $\AlgDiffF$ is recalled in \eqref{eq:fourierAlgDiff}. Thus, this numerical differentiation scheme falls within the framework of classical approaches for differentiation in the frequency domain, where a smoothing filter, in this case $\JacDiff$, is followed by an ideal differentiation operator $(\imag\omega)^n$. This property has first been observed in \cite{mboup2014}.

Analysing $\AlgDiffF$ is thus helpful to derive tuning guidelines by specifying desired frequency-domain properties. The approximation
\begin{equation}
\abs{\AlgDiffF(\omega)}\approx{\AlgDiffFTilde(\omega)}=\begin{cases}
1,&\quad \abs{\omega}\leq\AlgOmegaC,\\
\abs{\frac{\AlgOmegaC}{\omega}}^{\mu},&\quad \text{otherwise},
\end{cases}
\label{eq:lowPassFilterApp}
\end{equation}
of the amplitude spectrum of an algebraic differentiator has been proposed in \cite{kiltz2013}, where $\AlgOmegaC$ is defined in \eqref{eq:cutoffFreq} and $\mu=1+\min\{\alpha,\beta\}$. It follows that $\AlgDiffF$ can be interpreted as a low pass filter: The frequency $\AlgOmegaC$ and the set $\abs{\omega}<\AlgOmegaC$ are called cutoff frequency and passband, respectively. The set $\abs{\omega}>\AlgOmegaC$ is called stopband.

\begin{figure*}[h!]
	\centering
	\includegraphics{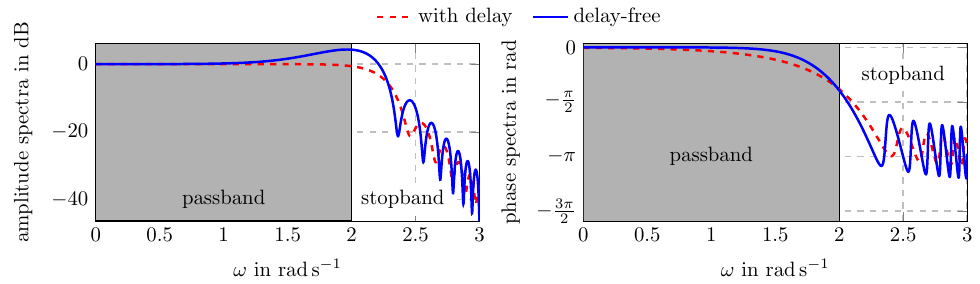}
	\caption{Amplitude and phase spectra of the differentiators with and without delay from \cref{list:frequency-domain-no-delay}.}
	\label{fig:spectra_no_delay}
\end{figure*}
Tuning guidelines proposed in \cite{kiltz2013,kiltz2017} can then be inferred from \eqref{eq:lowPassFilterApp}: A desired cutoff frequency $\AlgOmegaC$ and a desired filter order, i.e., the 	stopband slope given by $20\mu\,\SI{}{\dB}$, can be specified. Then, the filter window length   can be computed from \eqref{eq:cutoffFreq}.

It has been shown in \cite{mboup2018,othmane2021ejc,othmane2021b} that increasing $N$ increases the sensitivity to noise. It has also been shown in \cite{mboup2018} that the choice $\alpha=\beta$ yields differentiators with maximum robustness to noise. In addition, according to \cite{kiltz2017,othmane2021ejc}, and \cite{othmane2021b}, choosing $\alpha\neq\beta$ reduces stopband ripples in the amplitude spectrum.

\cref{exp:Specifying-frequency-domain} shows how a differentiator can be designed from desired frequency-domain properties. The computation of the amplitude and phase spectra is demonstrated. \cref{ex:delay} compares and discusses the amplitude spectra of differentiators with and without delay.
\begin{exmp}[Specifying frequency-domain properties]
	\label{exp:Specifying-frequency-domain}
	An algebraic differentiator with an amplitude spectrum asymptotically decreasing by $\SI{40}{\dB}$  per decade, i.e.,  $\min\{\alpha,\beta\}=1$,  and a cutoff frequency $\AlgOmegaC=\SI{100}{\radian\per\second}$ can be designed as described in \cref{list:frequency-domain}. The example also shows how the toolbox can be used to compute the amplitude spectrum, its approximation \eqref{eq:lowPassFilterApp}, and the resulting window length $T$. To increase the robustness with respect to corrupting noises the choices $N=0$ and $\alpha=\beta$ are made. Additionally, upper and lower bounds for $\AlgDiffF$ can be computed. They have been discussed in \cite{kiltz2017,othmane2021ejc,othmane2021b}.
	
	The amplitude spectrum, the approximation, the bounds, and the phase spectrum are given in \cref{fig:spectra}. Note that for this parametrisation the amplitude spectrum has infinitely many zeros. Thus, its lower bound is equal to zero (see also \cite{kiltz2017}).
	\qedsymbol
	\begin{center}
		\begin{lstlisting}[escapeinside={(*}{*)},frame=single,caption={Design of an algebraic differentiator from desired frequency-domain properties.}, label=list:frequency-domain,nolol]
# Create an instance of the class
diffA = AlgebraicDifferentiator(N=0,alpha=1.,beta=1.,T=None,wc=100)
# Compute the amplitude spectrum of the differentiator
omega = np.linspace(1,800,4*10**3)
ampA,phaseA = diffA.get_ampAndPhaseFilter(omega)
# Get the upper and lower bounds and the approximation of the amplitude spectrum
uA, lA, mA = diffA.get_asymptotesAmpFilter(omega)
# Get the filter window length
T = diffA.get_T()\end{lstlisting}
	\end{center}

\end{exmp}

\begin{exmp}[Differentiators with and without delay]
	\label{ex:delay}
	As discussed in \cref{sec:background}, algebraic differentiators can also be parametrised such that the estimated derivative is delay-free. This is done in \cref{list:frequency-domain-no-delay} for a differentiator with $N=1$ and $\alpha=\beta=1$. Its amplitude spectrum is also computed. For a comparison a differentiator with delay but the same parameters $\alpha$, $\beta$, and $N$ is also considered.
	
	\cref{fig:spectra_no_delay} shows the amplitude spectra of both differentiators from the latter example. The delay-free differentiator shows a significant overshoot. Thus, the resulting estimated derivative might not be usable. The decrease in the estimation quality of these differentiators has been remarked in \cite{mboup2009}. The overshoots in the amplitude spectra of delay-free differentiators have first been observed in \cite{mboup2014}.
	\qedsymbol

\end{exmp}

\begin{lstlisting}[escapeinside={(*}{*)},frame=single,caption={Comparison of amplitude spectra for differentiators with and without delay.}, label=list:frequency-domain-no-delay,nolol]
# Create an instance of the class for a differentiator with delay
diffA = AlgebraicDifferentiator(N=1,alpha=1.,beta=1.,T=None,wc=100)
# Create an instance of the class for a differentiator without delay
diffB = AlgebraicDifferentiator(N=1,alpha=1.,beta=1.,T=None,wc=100)
diffB.set_theta(1,False)

# Compute amplitude spectrum of the differentiator
omega = np.linspace(1,1000,8*10**2)
ampA,phaseA = diffA.get_ampAndPhaseFilter(omega)
ampB,phaseB = diffB.get_ampAndPhaseFilter(omega)\end{lstlisting}
\begin{figure*}[h!]
	\centering
	\includegraphics{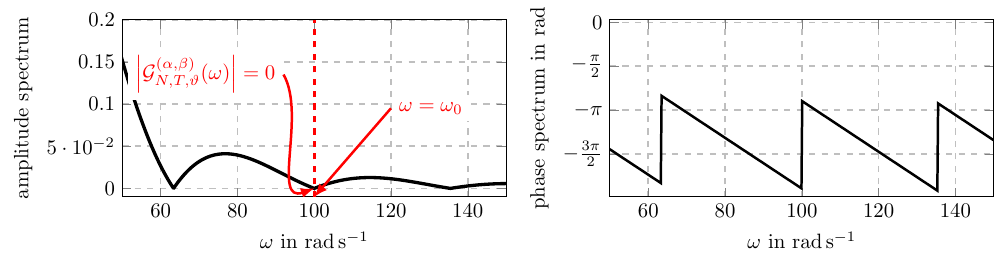}
	\caption{Amplitude and phase spectrum of the differentiator from \cref{list:frequency-domain-annihilation} for the annihilation of a harmonic disturbance with angular frequency $\omega_0$.}
	\label{fig:spectra_annihilation}
\end{figure*}

It has been observed in \cite{kiltz2013,kiltz2017} that the Fourier transform $\AlgDiffF$ of an algebraic differentiator with $N=0$ and $\alpha=\beta$ simplifies to 
\begin{equation*}
\mathcal{G}^{(\alpha,\alpha)}_{0,T,\vartheta}(\omega)=\mathrm{e}^{-\imag\tfrac{\omega T}{2}}\Gamma\left(\alpha+\frac{3}{2}\right)\left(\tfrac{4}{\omega T}\right)^{\alpha+\frac{1}{2}}\mathrm{J}_{\alpha+\frac{1}{2}}\left(\frac{\omega T}{2}\right),
\end{equation*}
where $\Gamma$ denotes the Gamma Function and $\mathrm{J}_{\alpha+\frac{1}{2}}$ is the Bessel function of the first kind and order $\alpha+1/2$. These special functions are defined in \cite{abramowitz2011}, for example. Thus, the amplitude spectrum exhibits an infinite number of positive zeros. Moreover, the differentiator provides phase linearity. By carefully choosing the parameters $\alpha$ and $T$ such that the zeros of $\mathcal{G}^{(\alpha,\alpha)}_{0,T,\vartheta}$ coincide with a specific frequency, the exact annihilation of disturbing harmonics can be performed. This parametrisation has been used in experimental studies in \cite{kiltz2012,kiltz2013,kiltz2013b,kiltz2014,kiltz2017}. Therein, the measured signals are corrupted by an unmodelled mechanical eigenmode of the system. Its influence on the control and estimation algorithms has been annihilated using the latter property. In \cite{kiltz2017}, this parametrisation has been used to approximate a matched filter. The following example demonstrates this special parametrisation.

\begin{exmp}[Exact annihilation of frequencies]
	Assume that the signal the derivative of which has to be estimated is corrupted by a harmonic signal of known frequency $\omega_0$. Choosing the window length $T$ such that
	$\omega_0T=2j_k$, with $j_k$ the $k$-th zero of the Bessel function of the first kind and order $\alpha+1/2$ yields $\mathcal{G}^{(\alpha,\alpha)}_{0,T,\vartheta}(\omega_0)=0$, i.e., the effects of the corrupting signal are exactly annihilated. This is demonstrated in \cref{list:frequency-domain-annihilation} for $\omega_0=\SI{100}{\radian\per\second}$ and $\alpha=2$. 
	\begin{center}
	\begin{minipage}{\linewidth}
\begin{lstlisting}[escapeinside={(*}{*)},frame=single,caption={Parametrisation of a differentiator to annihilate a known frequency.}, label=list:frequency-domain-annihilation,nolol]
# Create an instance of the class
w_0 = 100
alpha = 2.
# Choose the k-th zero of the Bessel function
k = 2
jk = float(mp.besseljzero(alpha+0.5,k))
T = 2*jk/w_0
diffA = AlgebraicDifferentiator(N=0,alpha=alpha,beta=alpha,
T=T)
# Evaluate the Fourier transform
omega = np.linspace(0,200,1000)
amp,phase = diffA.get_ampAndPhaseFilter(omega)\end{lstlisting}
\end{minipage}
	\end{center}

\cref{fig:spectra_annihilation} shows the amplitude and phase spectra of this differentiator and confirms the exact annihilation of the disturbing harmonic with frequency $\omega_0$. However, choosing an appropriate zero and a numerical value for $\alpha$ remain as the degrees of freedom for every specific application. The works \cite{kiltz2012,kiltz2013,kiltz2013b,kiltz2014,kiltz2017} have proposed an approach where these parameters are chosen such that a satisfying compromise between disturbance rejection and tolerable window length can be achieved. The size $T$ of the window is a crucial parameter since it influences the estimation delay, the computational burden, and the memory requirements of the implemented differentiators. Available hardware may imply constraints on the latter properties. \qedsymbol
\end{exmp}

\begin{figure*}[h!]
	\centering
	\includegraphics{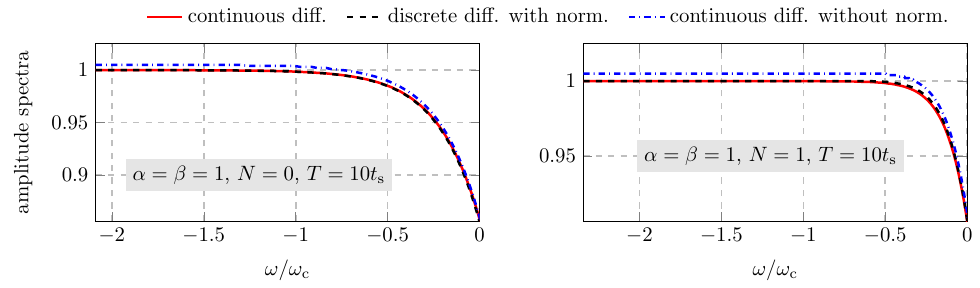}
	\caption{Amplitude of two continuous-time differentiators compared to those of discretised ones with and without normalisation.}
	\label{fig:effects_normalisation}
\end{figure*}

\subsection{Discrete-time implementation}

In most applications, the measured signal is available at discrete time instants only and the integral \eqref{eq:JacDiffApp} must be approximated by an appropriate quadrature method. Thus, the estimates are outputs of digital \acrshort{FIR} filters.

From the Nyquist–Shannon sampling theorem, it follows that when discretising a filter, the gain at the Nyquist frequency must be sufficiently small to reduce aliasing effects. As proposed in \cite{kiltz2017}, a possibility to avoid these effects is to specify the weakest relative attenuation
\begin{equation*}
k_{\mathrm{N,min}}=\frac{\omega_{\mathrm{N}}\abs{\AlgDiffF(\omega_{\mathrm{N}})}}{\AlgOmegaC\abs{\AlgDiffF(\AlgOmegaC)}}\approx\left(\frac{\AlgOmegaC}{\omega_{\mathrm{N}}}\right)^{\min\{\alpha,\beta\}+1-n_{\mathrm{max}}}
\label{eq:reltiveAttenation}
\end{equation*}
of an algebraic differentiator $\JacDiff$ at the Nyquist frequency $\omega_N = \pi/\samplingTime$, where $\samplingTime$ and $n_{\mathrm{max}}$ are the sampling period and the highest required derivative order, respectively (see also \cite{othmane2021b}).  \cref{ex:relative_attenuation} demonstrates the design of a differentiator with a desired relative attenuation $k_{\mathrm{N,min}}$.

\begin{exmp}[Specifying the relative attenuation $k_{\mathrm{N,min}}$]
	\label{ex:relative_attenuation}
	To achieve at least a desired relative attenuation of $k_{\mathrm{N,min}}$ for a given cutoff frequency $\AlgOmegaC$, the parameters $\alpha$ and $\beta$ have to be chosen such that
	\begin{equation*}
	\min\{\alpha,\beta\}=\frac{\ln\left(k_{\mathrm{N,min}}\right)}{\ln\left({\AlgOmegaC}/{\omega_{\mathrm{N}}}\right)}+n_{\mathrm{max}}-1.\label{eq:formulaAlpha}
	\end{equation*} 
	A differentiator is designed in \cref{list:relative_attenuation} for the approximation of a first order derivative with the cutoff frequency equal to $\SI{90}{\radian\per\second}$ and a relative attenuation equal to $10^{-3}=\SI{-30}{\dB}$. The considered sampling period is equal to $\SI{10}{\milli\second}$, the parameters $\alpha$ and $\beta$ are chosen equal, and $N=0$.  The resulting differentiator has a filter window length equal to $\SI{100}{\milli\second}$, i.e., $10$ sampling periods. The parameters $\alpha$ and $\beta$ are equal to $5.53$. The estimation delay is equal to $5$ sampling periods.
	\qedsymbol
	\begin{center}
		\begin{lstlisting}[escapeinside={(*}{*)},frame=single,caption={Design of a differentiator with a specified relative attenuation.}, label=list:relative_attenuation,nolol]
# Specify cutoff frequency
wc = 90
# Specify relative attenuation
kn = 10**(-3)
# Specify sampling period and Nyquist frequency
ts = 0.01
wN = np.pi/ts
# Specify order of derivative
der_order = 1
# Compuate parameter (*${\color{codegreen}\alpha}$*)
alpha = np.log(kn)/np.log(wc/wN)+der_order-1
# Create an instance of the class with delay
diffA = AlgebraicDifferentiator(N=0,alpha=alpha,beta=alpha,
T=None,wc=wc,ts=ts)
# Get window length
T = diffA.get_T()
# Get delay
delay = diffA.get_delay()\end{lstlisting}
	\end{center}
\end{exmp}

In the following, it is assumed that the window length $T$ is an integral multiple of the sampling period, i.e., $\JacT=\AlgTdis\samplingTime$, and equidistant sampling is considered for simplicity. The abbreviation $y_k = y(k \samplingTime)$, $k\in\setN$, for a sample of a signal $y$ at time $k \samplingTime$ is used. Then, a  discrete-time approximation 
\begin{equation}
\hat{y}^{(n)}_{k+\theta}=\frac{1}{\Phi}\sum_{i=0}^{L-1}w_iy_{k-i},\quad \Phi=\frac{\samplingTime^n}{n!}\sum_{k=0}^{L-1}w_k(-k)^n,
\label{eq:filterDiscrete}
\end{equation}
of \eqref{eq:JacDiffApp} can be achieved. Therein, $\theta$, $L$, and $w_i$ depend on the numerical integration method used as discussed in \cite{kiltz2017,othmane2021b,othmane2022thesis}. For example, the mid-point rule yields $\theta=\frac{1}{2}$, $w_i=\samplingTime g^{(n)}_{i+\theta}$, and $L=\AlgTdis$.
The normalisation factor $\Phi$ has been first introduced in \cite{kiltz2017} to guarantee that the DC component of the sought derivative is preserved. The effects of normalisation are demonstrated in the following example using the functions of the toolbox.
\begin{exmp}[Effect of the normalisation factor]
	Per default, the toolbox implements the normalisation presented in \eqref{eq:filterDiscrete}. The code in \cref{list:normalization} demonstrates the effects of the normalisation by comparing the amplitude spectra of two differentiators: The first one is discretised with the normalisation and the second is without.

	\begin{center}
		\begin{lstlisting}[escapeinside={(*}{*)},frame=single,caption={Discretisation of differentiators with and without normalisation.}, label=list:normalization,nolol]
# Create an instance of the class for a differentiator with the normalisation
ts = 0.01
diffA = AlgebraicDifferentiator(N=1,alpha=1.,beta=1.,
T=10*ts,ts=ts)
# Create an instance of the class for a differentiator without the normalisation
diffN = AlgebraicDifferentiator(N=1,alpha=1.,beta=1.,
T=10*ts,ts=ts,corr=False)
# Evaluate Fourier transform of the continuous-time and discrete-time differentiators for comparisons
omega = np.linspace(0,200,1000)
method = 'mid-point'
d = 0
ampA_C,phaseA_C = diffA.get_ampAndPhaseFilter(omega)
ampA_D,phaseA_D = diffA.get_ampSpectrumDiscreteFilter(omega, d, method=method)
ampN_D,phaseN_D = diffN.get_ampSpectrumDiscreteFilter(omega, d, method=method) \end{lstlisting}
	\end{center}
\cref{fig:effects_normalisation} shows the spectra of two continuous-time differentiators discretised with and without the normalisation factor. It can be seen that the DC component is only preserved when the normalisation is used. 
\qedsymbol
\end{exmp}
\begin{figure*}[h!]
	\centering
	\includegraphics{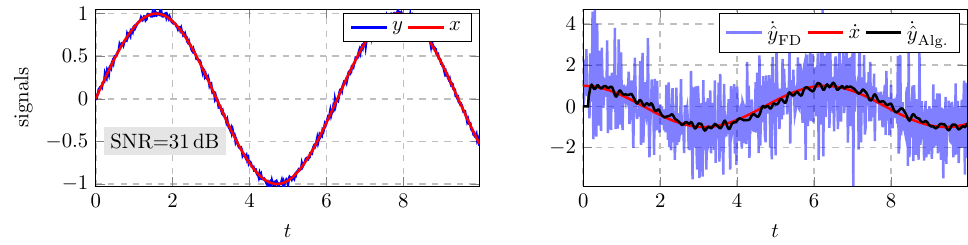}
	\caption{Estimation of the first order derivative $\dot{x}$ of a sinusoidal signal $x$ using a measurement $y$ using the forward-difference method (FD) and the algebraic differentiator (Alg.) designed in \cref{list:estimation}.}
	\label{fig:estimated_der}
\end{figure*}

Algebraic differentiators and their tuning guidelines have been discussed in the continuous-time domain. Thus, the discretised ones have to preserve the specified properties. Since most tuning guidelines discussed here rely on the frequency-domain interpretation of the differentiators, the analyses of the discretisation effects can be performed by comparing the Fourier transform of the continuous and discrete differentiators. This can be done using the cost function
\begin{equation}
	\mathcal{J}=\frac{\int_{0}^{\Omega}\abs{\mathcal{F}\left\{g^{(n)}-\hat{g} ^{(n)}\right\}(\omega)}^2\diff\omega}{\int_{0}^{\Omega}\abs{\mathcal{F}\left\{g^{(n)}\right\}(\omega)}^2\diff\omega}
	\label{eq:cost}
\end{equation}
introduced in \cite{kiltz2017} (see also \cite{othmane2021b,othmane2022thesis}), where $\Omega$ has to be chosen according to the frequency interval of interest. In the latter cost function,  $\omega\mapsto{\mathcal{F}\left\{g^{(n)}\right\}(\omega)}$ is the Fourier transform of the continuous-time differentiator $g=\JacDiff$. The discrete-time filter from \eqref{eq:filterDiscrete} is denoted $\hat{g}^{(n)}$ and its Fourier transform is
\begin{equation}
\omega\mapsto{\mathcal{F}\left\{\hat{g}^{(n)}\right\}(\omega)}={\frac{1}{\Phi}\sum_{k=0}^{L-1}w_k\mathrm{e}^{-\imag\omega(k+\theta)\samplingTime}}.
\end{equation}
The cost function $\mathcal{J}$ compares the discretisation error with the noise amplification of the continuous-time differentiator (the reader is referred to \cite{othmane2021b,kiltz2017} for further details). Thus, it can be used to decide if the former can be neglected compared to the latter. The cost function can be easily computed using the functions in the toolbox as demonstrated in \cref{ex:discretisation_error}.

\begin{exmp}[Discretisation error]
	\label{ex:discretisation_error}
	The cost function \eqref{eq:cost} is computed in \cref{list:discretization_error} for a differentiator with the parameters $\alpha=\beta=7$, $N=0$, and $T=20\samplingTime$. 
	
	\begin{center}
		\begin{lstlisting}[escapeinside={(*}{*)},frame=single,caption={Evaluation of the discretisation error.}, label=list:discretization_error,nolol]
# Create an instance of the class for a differentiator with delay
ts = 0.001
diffA = AlgebraicDifferentiator(N=0,alpha=7.,beta=7.,
T=20*ts,ts=ts)
# Compute the discretization error
Omega = np.pi/ts
J = diffA.get_discretizationError(1,Omega)\end{lstlisting}
	\end{center}
The resulting cost $\mathcal{J}$ is approximately equal to $1.6\cdot10^{-10}$. Thus, the discretisation effects can be neglected.  The variation of $\mathcal{J}$ with respect to the parameters of the differentiator has been analysed in \cite{kiltz2017,othmane2022thesis}, for example. \qedsymbol
\end{exmp}

\subsection{Estimation of derivatives using AlgDiff}
\cref{ex:numerical_differentiation} demonstrates the numerical differentiation of a given signal using the implemented functions and a differentiator with desired frequency-domain properties.
\begin{exmp}[Numerical differentiation]
	\label{ex:numerical_differentiation}
	An algebraic differentiator with the cutoff frequency ${\AlgOmegaC=\SI{20}{\radian\per\second}}$ and a desired relative attenuation $k_{\mathrm{N},\min}=10^{-3}$ shall be designed for the estimation of the first order derivative of a signal sampled with a sampling period equal to $\SI{20}{\milli\second}$. As in \cref{exp:Specifying-frequency-domain}, the choice $N=0$ and $\alpha=\beta$ is made to increase the robustness with respect to disturbances. The resulting differentiator has the parameters	 $\alpha=\beta=3.35$ and  a window length equal to $14\samplingTime$. The estimation delay corresponds to $7\samplingTime$. The cost function $\mathcal{J}$ is equal to $7.9\cdot10^{-6}$. Thus, discretisation effects can be neglected. The parametrisation of the differentiator and the estimation of the derivative of a noisy signal $t\mapsto y(t)=\sin(t)+\eta(t)$ is shown in \cref{list:estimation}. The additive disturbance $\eta$ is drawn from a Gaussian distribution with standard deviation equal to $0.02$ such that the signal to noise ratio is $\text{SNR}=\mathrm{log}_{10}\left(\tfrac{\sum_{i}y^2(t_i)}{\sum_{i}\eta^2(t_i)}\right)\approx\SI{31}{\dB}$.

The estimation results are shown in \cref{fig:estimated_der} and compared to the results of a simple forward difference method. For a comparison with more advanced methods the reader is referred to \cite{othmane2021ejc,sidhom2011,liu2011}. This example shows the advantages of algebraic differentiators and their systematic tuning guidelines. Furthermore, it highlights the user friendliness of the toolbox AlgDiff. \qedsymbol	
	\begin{center}
		\begin{lstlisting}[escapeinside={(*}{*)},frame=single,caption={Approximation of a derivative of a given signal using a differentiator designed with desired frequency-domain properties.}, label=list:estimation,nolol]
# Design differentiator
ts = 0.02
kn = 10**(-3)
wN = np.pi/ts
a = np.log(kn)/np.log(wc/wN)
diffA = AlgebraicDifferentiator(N=0, alpha=a, beta=a, T=None, ts=ts,wc=20)
J = diffA.get_discretizationError(1,np.pi/ts)
# Define the signal
t = np.arange(0,10,ts)
a0 = 1
w0 = 1
x = a0*np.sin(w0*t)
dx = a0*w0*np.cos(w0*t)
eta = np.random.normal(0,0.02,len(t))
y = x+eta
# Estimate derivative
dyApp = diffA.estimateDer(1,y)\end{lstlisting}
	\end{center}

\end{exmp}

\section{Conclusion}
This contribution discusses algebraic differentiators and the use  of the toolbox AlgDiff. Properties of the differentiators are recalled and the use of the functions implemented in AlgDiff was shown. The design of differentiators with desired frequency-domain properties and the analysis of discretisation issues are addressed with AlgDiff and the results are discussed and illustrated. It is shown that with few lines of code, efficient differentiators can be designed and used for the numerical differentiation of signals. Further works where the systematic tuning of the differentiators are discussed in detail are \cite{kiltz2013,kiltz2014,kiltz2013b,kiltz2017,othmane2021ejc,othmane2021b,othmane2022thesis} and \cite{scherer2023}, for example. An extensive list of applications and more examples for the tuning of the differentiators are provided in \cite{othmane2021b} .
\label{sec:conclusion}

\printbibliography

\begin{appendices}
	\renewcommand{\theequation}{A.\arabic{equation}}
\section{Jacobi polynomials}
\label{sec:app1}
The Jacobi polynomial $\jacPol_{N}^{(\Jacalpha, \Jacbeta)}$ can be defined as
\begin{align*}
\jacPol_{N}^{(\Jacalpha, \Jacbeta)}(\tau) &= \sum_{k=0}^{N} c_{k}^{(\Jacalpha, \Jacbeta)} (\tau - 1)^{k},\\
c_{k}^{(\Jacalpha, \Jacbeta)} &= \tfrac{\JacGamma(\Jacalpha + N + 1) \JacGamma(\Jacalpha + \JacTheta + N + k + 1)}{2^{k}N! \JacGamma(\Jacalpha + \Jacbeta + N + 1) \JacGamma(\Jacalpha + k + 1)},
\end{align*}
where $\tau \in\mathcal{I}_P=[-1,1]$, $\JacN\in \mathbb{N}$, $\alpha,\beta>-1$, and $\JacGamma$ is the Gamma function (see \cite[Sections~6.1 \& 22.2]{abramowitz2011} for the definitions). The polynomials are orthogonal in the interval $\mathcal{I}_P$ with respect to the weight function
\begin{align}
\wAlphaBeta(\tau) & = \begin{cases}
(1 - \tau)^{\Jacalpha} (1 + \tau)^{\Jacbeta}, & \tau \in \mathcal{I}_P,\\
0, & \text{otherwise}.
\end{cases}
\label{eq:weight}
\end{align}
\section{Kernel of algebraic differentiators}
\label{sec:app2}
The kernel $\JacDiff$ in \eqref{eq:JacDiffApp} is defined as
\begin{align}
\JacContiKernel{t} & = \tfrac{2}{T}\sum_{k=0}^{N}\tfrac{\jacPol_{k}^{(\alpha,\beta)}(\vartheta)}{\norm{\jacPol_{k}^{(\alpha,\beta)}}^2}\left(w^{(\alpha,\beta)}\cdot\jacPol_{k}^{(\alpha,\beta)}\right)\circ\!\left(\theta_{T}(\tau)\right),
\label{eq:JacDiffKernel}
\end{align}
with $\theta_{T}(t) = 1 - 2t/\JacT$ and $\norm{x} = \sqrt{\SkalarProdukt{x, x}}$ the norm induced by the inner product $${\SkalarProdukt{x,y} = \int_{-1}^{1} \wAlphaBeta(\tau) x(\tau) y(\tau) \diffd \tau}.$$

\section{Impulse and step responses}
\label{sec:app3}
\label{sec:app:responses}
The impulse and step responses of an algebraic differentiator $\JacDiff$ are by \eqref{eq:JacDiffKernel}
 and\begin{subequations}
	\begin{align*}
	\stepJac(\tau) &= \begin{cases}
	0,&\;\text{for}\; \tau<0\\
	\regBetaFcn{\tfrac{\tau}{T}}(\bar\alpha,\bar\beta)+f_{N,T,\vartheta}^{(\alpha,\beta)}(\tau),&\;\text{for}\; \tau\in[0,T]\\
	1&\;\text{otherwise},
	\end{cases},
	\end{align*}
\end{subequations}
respectively, with $\bar\alpha=\alpha+1$, $\bar\beta=1+\beta$, $\regBetaFcn{\tau}$ the regularized incomplete Beta function defined in \cite[Section~6.6]{abramowitz2011}, and
\begin{equation*}
f_{N,T,\vartheta}^{(\alpha,\beta)}(\tau) = \sum_{k=1}^{N}\tfrac{\jacPol_{k}^{(\alpha,\beta)}(\vartheta)}{2k\norm{\jacPol_{k}^{(\alpha,\beta)}}^2}\left(w^{(\bar\alpha,\bar\beta)}\cdot\jacPol_{k-1}^{(\bar\alpha,\bar\beta)}\right)\circ\!\left(\theta_{T}(\tau)\right).
\end{equation*}
\section{Fourier transform $\AlgDiffF$}
\label{sec:app4}
Closed forms of the Fourier transform $\AlgDiffF$ of an algebraic differentiator $\JacDiff$ have been derived in \cite{kiltz2012,kiltz2017} and \cite{mboup2018}. One possible representation of $\AlgDiffF$ is
\begin{equation}
	\begin{aligned}
		\AlgDiffF(\omega)&=\sum_{i=0}^{N}\tfrac{(\alpha+\beta+2i+1)\jacPol_{i}^{(\alpha,\beta)}(\theta)}{\alpha+\beta+i+1}\times\\&\sum_{k=0}^{i}(-1)^{i-k}\binom{i}{k}\mathrm{M}_{i,k}^{(\alpha,\beta)}(-\imag\omega T),\quad \imag^2=-1,
		\label{eq:fourierAlgDiff}
	\end{aligned}
\end{equation}
with 
\begin{equation}
\mathrm{M}_{i,k}^{(\alpha,\beta)}(z)=\mathrm{M}(\alpha+i-k+1,\alpha+\beta+i+2,z),
\end{equation}
where $z\in\mathbb{C}$ and $\mathrm{M}$ is the confluent hypergeometric function defined in \cite[Section~13.1]{abramowitz2011}.

\section{Cutoff frequency $\AlgOmegaC$}
\label{sec:app5}
Let $\mu=1+\min\{\alpha,\beta\}$, $\kappa=\abs{\alpha-\beta}$. The cutoff frequency of an algebraic differentiator is
\begin{align}
\AlgOmegaC&=\frac{1}{T}\left(\frac{q_{N,\vartheta}^{(\alpha,\beta,\sigma)}}{\Gamma(\mu+\kappa)}\right)^{\frac{1}{\mu}}\label{eq:cutoffFreq},
\end{align}
where
\begin{align*}
q_{N,\vartheta}^{(\alpha,\beta,\sigma)}&=\begin{cases}
\Gamma(\mu)\max\left\{\abs{r_{N,T}^{(\mu,0,\sigma)}},s_{N,T}^{(\mu,0,\sigma)}\right\},&\quad \kappa=0,\\
\Gamma(\mu+\kappa)\abs{r_{N,T}^{(\mu,\kappa,\sigma)}},&\quad\kappa>0,
\end{cases}\\
\sigma&=\begin{cases}
1,&\alpha\leq\beta,\\
-1,&\alpha>\beta,
\end{cases}\\
r^{(\mu,\kappa,\sigma)}_{N,\vartheta}
&=\sum_{i=0}^{N}\frac{c_i^{(\mu,\kappa)}}{\Gamma(\mu+\kappa+i)}\jacPol_{i}^{(\mu-1,\mu+\kappa-1)}(\sigma\vartheta),\\
s^{(\mu,\kappa,\sigma)}_{N,\vartheta}
&=\sum_{i=0}^{N}(-1)^i\frac{c_i^{(\mu,\kappa)}}{\Gamma(\mu+i)}\jacPol_{i}^{(\mu-1,\mu+\kappa-1)}(\sigma\vartheta),\\
c_i^{(\mu,\kappa)}&=\left(2\mu+\kappa+2i-1\right)\Gamma(2\mu+\kappa+i-1).
\end{align*}

\end{appendices}

 \begin{contributors}
	
	     \contributor{Amine Othmane}
	         {Systems Modeling and Simulation, Saarland University, Saarbrücken, Germany}
	         {amine.othmane@uni-saarland.de}
	         {}
	         {Amine Othmane received a bachelor’s degree in Mechatronik from Saarland University and a master’s degree in Engineering Cybernetics from the University of Stuttgart. He has been a visiting student at the University of Alberta in Edmonton, Canada, and the Norwegian University of Science and Technology in Trondheim, Norway. He received his doctoral degree in 2022 from Saarland University and Université Paris-Saclay, France, in the context of a joint international doctoral supervision (“cotutelle''). Since 2023 he works with Prof. Kathrin Flaßkamp (Systems Modeling and Simulation, Saarland University) on combining physics-based and data-based approaches for wind energy systems. He received the best presentation award at the 57. Regelungstechnisches Kolloquium in Boppard. His main research topics are numerical differentiation, parameter estimation, fault diagnosis, model-free control, numerical methods, and physics-based AI methods.}
	
	     \contributor{Joachim Rudolph}
	         {Chair of Systems Theory and Control Engineering, Saarland University, Germany}
	         {j.rudolph@lsr.uni-saarland.de}
	         {}
	         {Joachim Rudolph received the doctorate degree from Université Paris XI, Orsay, France, in 1991, and the Dr.-Ing.\ habil.\ degree from Technische Universität Dresden, Germany, in 2003. Since 2009, he has been the Head of the Chair of Systems Theory and Control Engineering at Saarland University, Saarbrücken, Germany. His current research interests include controller and observer design for non-linear and infinite dimensional systems, algebraic systems theory, and the solution of demanding practical control problems.}
	
	 \end{contributors}

\end{document}